\documentclass{ecai}
\usepackage{times}
\usepackage{graphicx}
\usepackage{latexsym}
\usepackage[hyphens]{url}
\usepackage{hyperref}
\hypersetup{breaklinks=true}
\usepackage[table,xcdraw]{xcolor}
\usepackage{booktabs,chemformula}
\usepackage{longtable}
\usepackage{lscape}
\usepackage[export]{adjustbox}

\newcommand{\eg}{\emph{e.g.}, }       
\newcommand{\ie}{\emph{i.e.}, }      

\begin{document}

\title{Progressing Towards Responsible AI}

\author{Teresa Scantamburlo \institute{European Centre for Living Technologies, Ca' Foscari University of Venice, Italy, email: teresa.scantamburlo@unive.it} \and Atia Cort\'es \institute{Barcelona Supercomputing Center, Spain, email: atia.cortes@bsc.es} \and Marie Schacht \institute{QuantumBlack, UK, email: Marie.Schacht@quantumblack.com}}
\maketitle
\bibliographystyle{ecai}

\begin{abstract}
The field of Artificial Intelligence (AI) and, in particular, the Machine Learning area, counts on a wide range of performance metrics and benchmark data sets to assess the problem-solving effectiveness of its solutions. However, the appearance of research centres, projects or institutions addressing AI solutions from a multidisciplinary and multi-stakeholder perspective suggests a new approach to assessment comprising ethical guidelines, reports or tools and frameworks to help both academia and business to move towards a responsible conceptualisation of AI.
They all highlight the relevance of three key aspects: \textit{(i)} enhancing cooperation among the different stakeholders involved in the design, deployment and use of AI; \textit{(ii)} promoting multidisciplinary dialogue, including different domains of expertise in this process; and \textit{(iii)} fostering public engagement to maximise a trusted relation with new technologies and practitioners. 
In this paper, we introduce the Observatory on Society and Artificial Intelligence (OSAI), an initiative grew out of the project AI4EU aimed at stimulating reflection on a broad spectrum of issues of AI (ethical, legal, social, economic and cultural). In particular, we describe our work in progress around OSAI and suggest how this and similar initiatives can promote a wider appraisal of progress in AI.
This will give us the opportunity to present our vision and our \emph{modus operandi} to enhance the implementation of these three fundamental dimensions.
\end{abstract}

\section{Introduction}

In science, nothing has been more controversial than the notion of progress. Debates on what progress is and if it does really exist abound in the philosophy of science and are closely related to questions about the goals and the methods of a scientific discipline \cite{sep-sci-prog}. Although between the 1960s and 1970s philosophers of science put forward thought-provoking views, the idea of progress is commonly associated with the incremental acquisition of knowledge in a particular domain. The same idea is still prevalent in toady's research practice, including Artificial Intelligence (AI). For instance, a study discussing the flaws in Google Flu Trends claimed that \textit{``science is a cumulative endeavour, and to stand on the shoulders of giants requires that scientists be able to continually assess work on which they are building''} \cite[p 1205]{lazer14}.

In this paper, we commit to the view that the progress of a scientific discipline can be measured by the problem-solving\footnote{The notion of problem-solving also describes a specific AI research program (\eg the General Problem Solver \cite{newell59}). However, in this paper, this idea refers to the problem-oriented activity of science as neatly described by \cite{lau78}.} effectiveness of its theories \cite{lau78}. This view applies to, not only science, but also to different intellectual endeavours, including areas where solutions consist of technical artefacts such as algorithms and computing systems. 

Typically, the problem-solving effectiveness of AI solutions is a matter of performance testing. For example, in supervised learning, we assess algorithms based on the number of errors that they make on new, unseen data. Performance lies at the very heart of any learning algorithm, which is, by definition, a computer program that improves through experience. 

Intuitively, a Machine Learning (ML) technique is progressive as it performs well along with distinct criteria, including the task at hand, the benchmark data set and the computed performance measure. 
Although there is no hint of absolute progress in the field - in that no algorithm has proved to be the best at any possible task or condition -, deep learning methods have nevertheless hit the mark in multiple domains ranging from diagnosing eye diseases \cite{defauw18} to playing game \cite{silver17}. Other signs of progress regard the time of processing \cite{aindex19} which relies on the evolution of CPU capabilities. 


However, the introduction of AI algorithms into large portions of human life has suggested that technical performance is not enough. The adequacy of AI solutions depends on a broader set of considerations accounting for the behaviour of AI systems within the environment they are embedded. 
Repeated facts of algorithmic discrimination and lack of transparency shifted the focus from performance to accountability, from advances in accuracy and speed of computation to the protection of human rights and democratic values. In other words, the appraisal of AI progress is moving away from a purely technical assessment and becoming a multi-factorial affair which integrates aspects of privacy, fairness and transparency, among others. 

The transition towards a broader notion of AI assessment is the main focus of the present work. As we will see, a new \emph{ethical turn} prompted the rise of centres and networks addressing AI solutions from a multidisciplinary and multi-stakeholder perspective. 


Our maim claim is that these initiatives have contributed to moving towards a different notion of progress in AI that goes beyond technical performance to foster responsible development and dissemination of AI, acting on three main elements: \textit{(i)} promoting an interdisciplinary dialogue around AI; \textit{(ii)} involving diverse stakeholders and \textit{(iii)} engaging the public. In particular, we would like to present the Observatory on Society and AI as an example of these multidisciplinary and multi-stakeholder initiatives. 

The paper is structured as follows. In Section \ref{sec:assess}, we describe traditional practices for the assessment of AI, focusing in particular on ML, and the attempts to change them. Section \ref{sec:osai} introduces the Observatory for Society and AI as an example of an initiative which may contribute to the \emph{ethical turn} of AI. In Section \ref{sec:WGs}, we survey some responsible practices which will be part of the inventory of resources to be explored by the Observatory. We will conclude in Section \ref{sec:conl} with some final remarks.





\section{The assessment of progress in AI} \label{sec:assess}

One of the keys to understanding the progress of a scientific discipline is to assess its ability to solve problems. Disciplines have developed several methodologies to assess their own problem-solving effectiveness, but what it means for a problem to be solved can vary across the fields. 
For example, in logic, a problem can be viewed as solved when a theorem has been proved, 
in philosophy when a thesis is well-argued, in medicine when a treatment cures a disease, and so forth. Each discipline may develop different standards and processes for accepting solutions. The efficacy of a philosophical argument is thus assessed differently from the efficacy of an experimental result.

The field of ML has developed various practices to assess the effectiveness of learning techniques. Most of them reflect the ideal of experimentation in natural science, a model recommended particularly in the early days of ML research \cite{lang88}, then taken as a sign of maturity and objectivity of the field \cite{drumm06}. Usually, testing ML algorithms involve the use of benchmark data sets, the selection of performance measures, multiple tests and comparisons with competing methods. 
In a classification task, for example, a typical performance metric is accuracy, a scalar value which represents the fraction of predictions that an algorithm got right. 
More sophisticated methods look at the optimal trade-off between benefits (the true positive rate) and costs (the false-negative rate) such as ROC analysis.

The choice of the performance measure is a crucial task as the available metrics have a distinct meaning which depends on the context of an application. In other words, the measured value represents something we care about \cite{drumm06}. 
For example, in a system predicting fraud attempts by loan applicants, testing for precision might be sufficient and more informative than other metrics such as specificity. Things would change if the outcome referred to cancer detection, where misclassification comes at different cost. 

The creation and maintenance of large data repositories is another influential component of ML testing practice. Since the appearance of UCI collection \cite{UCI}
, the field has mostly committed to a benchmark-oriented attitude where data sets from disparate domains become the reference point for comparing algorithms' performance. In recent times, the activity of data collection has witnessed a massive surge and new large-scale databases, as well as more productive data annotation practices, have come to the surface. 

The ImageNet project \cite{imNet} generated more than 14 million images annotated by thousands of crowdworkers and structured around the WordNet hierarchy \cite{wNet}
. ImageNet has also inspired contests (the so-called ImageNet Large Scale Visual Recognition Challenge\footnote{See: \url{http://www.image-net.org/challenges/LSVRC/}}) where ML practitioners can compete and test their models in different specific tasks, such as object detection and image segmentation. Benchmarks and competitions abound also in natural language processing where we count challenges for question-answering, reasoning and sentiment analysis\footnote{See \eg the General Language Understanding Evaluation (GLUE) benchmark (\url{https://gluebenchmark.com/}, the the Stanford Question Answering Dataset (SQuAD) challenge \url{https://rajpurkar.github.io/SQuAD-explorer/} and the competition section within the CodaLab platform \url{https://codalab.org/}}.  

In recent times, the assessment of progress in AI was felicitated by initiatives tracking algorithms' performance during competitions, open repositories and code platforms. 
These include, for example, the AI index initiative \cite{aindex19} and the AI Watch methodology for the monitoring of AI progress \cite{AiWatchMet}. Thanks to them, one can get a glimpse of the significant breakthroughs achieved by the field. For example, in large-scale object classification tasks, the classification error of best-performing algorithms fell from 0.28 to 0.023 (see the results presented at CVPR Workshop 2017).  

While the systematic analysis of technical performance tells us that the field is progressing at a fast pace, there were AI researchers casting doubts on the robustness of standard assessment approach and claiming that, in reality, what we call progress could be only an illusion \cite{hand06}. 





\subsection{The ethical turn}

The discontent with standard testing practice in ML research is not a recent phenomenon. As early as 2006 Chris Drummond \cite{drumm06} criticised the standard testing approach raising three key points:
\begin{enumerate}
  \item \emph{Performance measures}: he observes that there are other factors influencing a performance measure like accuracy and these may include misclassification costs, the stability of the error rate, and the needs of the end users, among others.
  \item \emph{Statistical tests}: following criticism in Psychology, he highlights that statistical tests are frequently misinterpreted, for example, as a confirmation of the alternative to the null hypothesis\footnote{In statistics the notion of the null hypothesis refers to a default assumption such as \textit{``data has a normal distribution''} or \textit{``there is no correlation between two random variables''}. In short, a statistical test tells us if we should reject or fail to reject the null hypothesis. In ML the null hypothesis assumes that there is no difference between the performance measures of two algorithms. The one with the greater value is the winner algorithm.}. While their employment in ML experimental practice is often taken as a sign of rigor and objectivity, he contends that they do not give the degree of evidence that many people believe.
  \item \emph{Benchmark data sets}: while the use of benchmark data sets allow us to easily compare algorithms' performance, they suffer from serious limitations. For example, drawing on previous analysis, he raises concerns as to whether benchmark data sets are really representative of reality and how much the data collection and construction account for differences in class distribution.
\end{enumerate}
 
Further impulse to renovate the experimental practice in ML came from the movement of reproducible research requiring the publication of code and data to reproduce the results reported in scientific articles\cite{sonn07}. In a similar spirit, important ML conferences and journals encourage authors to adhere to best practices by making available data and software tools\footnote{See \eg the eligibility criteria in the call for paper of the European Conference on Machine Learning (ECML). The Journal for Artificial Intelligence Research has even stricter requirements.}. Other relevant initiatives include open source repositories and platforms allowing AI practitioners to share data and AI models, such as Papers with Code \cite{PwC} and Open ML \cite{OML}. Another recent work proposed a more comprehensive view of assessment integrating classical performance measures with often neglected costs connected to the development and deployment of an AI system \cite{Plumed18}. 

While these efforts operated within the edges of traditional scientific principles such as transparency and reproducibility, other assessment criteria emerged as a consequence of numerous debates around the impact of AI on society along with plans for action, what we call the \emph{ethical turn} of AI. The new wave of optimism prompted by the successful application of AI to big data \cite{hale09} was followed by critical analyses raising issues for culture \cite{boyd12, kit14} and human decision-making \cite{baro16}. Other failures and overstatements in the application of AI to transport and healthcare stimulated many initiatives around the world. These comprise research projects, journalistic investigation, centres and networks focused on the impact of AI on our lives.

In the few last years, we have witnessed more than one hundred declarations of AI principles from governments, organizations and multi-stakeholder initiatives, aimed at providing normative guidance for ethical, rights-respecting, and socially beneficial development and use of AI technologies. Most of those guidelines are aligned on the following eight key themes: Privacy, Accountability, Safety and Security, Transparency and Explainability, Fairness and Non-discrimination, Human Control of Technology, Professional Responsibility, and Promotion of Human Values \cite{Fjeld20}.

Note that, while ethical considerations are not new in the field of AI and notable scholars, such as Norbert Wiener, had already warned of the possible misuse of intelligent and control systems \cite{Wiener}, the scale and the number of present efforts have no precedents in the history of the field. For this reason, we acknowledge all these initiatives as a whole movement with the potential to widen future assessment practices. 

A final, interesting remark regards the variety of actors who are contributing to this ethical wave. The European Commission, for instance, based on the Trustworthy AI Guidelines (TAIG) \cite{HLEG2019} published by the High Level expert Group on AI (HLEG-AI), proposed a regulatory framework for high-risk AI applications with a view to build an ``ecosystem of trust'' \cite{white20}. Big companies, likewise, published new design principles\footnote{See \eg the AI Guidelines of Telefónica \url{https://www.telefonica.com/en/web/responsible-business/our-commitments/ai-principles} or Deutsche Telekom \url{https://www.telekom.com/en/company/digital-responsibility/details/artificial-intelligence-ai-guideline-524366}} and audit frameworks \cite{accuntgap}, also motivated by practical needs which are usually attenuated in a research context (think of company's liability and reputation). Finally, the landscape of ethical activities comprises a large number of centres (see Table \ref{tab:centres}) which scrutinize AI systems through the lens of legal principles and social values, study the impact on economy and human labour, and engage lay people with educational material or works of art.

\section{An Observatory for Society and AI (OSAI)} \label{sec:osai}

The Observatory on Society and Artificial Intelligence (OSAI or simply ``Observatory'' hereafter) was set up in 2019 within the H2020 EU funded project AI4EU \cite{AI4EU}, whose objective is to build the first European AI on-demand web platform and ecosystem. The OSAI is an example of this vast array of initiatives animating the \emph{ethical turn} of AI outlined in the previous section. Though at its infancy, it gives us the opportunity to explore how this and similar activities can contribute to stretch the assessment of AI and turns progress towards ethical principles.

The OSAI's aim is to support discussion and to facilitate the distribution of information about the Ethical, Legal, Socio-Economic and Cultural issues of AI (ELSEC-AI) within Europe. Specifically, the OSAI has the following objectives:
\begin{itemize}
    \item To stimulate reflection, discussion and due consideration of ELSEC-AI issues within the project through a series of working groups (see Section \ref{sec:WGs}). OSAI is attracting a network of experts in different domains of ELSEC-AI that will contribute to bridge the knowledge gap existing today within AI practitioners and users.
    \item To provide resources to educate the general EU public more accurately about AI and ELSEC-AI issues by generating weekly content in the form of articles, reports, cultural announcements with the objective to promote discussion and awareness on these topics.
\end{itemize}

The Observatory evolves in a complex scenario: the field of AI is gaining momentum, and many public and private agencies have begun to consider the opportunities and the risks that lie behind this exciting trend. The OSAI seeks to carve out its own identity and role neither in contrast nor competition with other existing European initiatives (\eg HLEG-AI). It aims to increase connections among these related projects and make accessible a broad range of articles to the European public at large. 
The OSAI's approach can be described by three verbs: \textit{1) Observe} facts and events occurring within Europe by monitoring newspapers, online bulletins, scientific literature, etc.; \textit{2) Reflect} on particular events or issues through to the contribution of ELSEC-AI experts and, in particular, thanks to the activities of the working groups; \textit{3) Report} to the general public by using a simple (but not simplistic) language in a way to support mutual understanding among experts and educate lay people.



\subsection{The context}

\begin{table*}[]
\centering
\caption{European centres for AI and Ethics}
\label{tab:centres}
\begin{tabular}{|p{0.15\textwidth}|p{0.1\textwidth}|p{0.15\textwidth}|p{0.5\textwidth}|}
\hline
\textbf{Name} &
  \textbf{Country} & \textbf{Type} &
  \textbf{Objective} \\ \hline
HumanE AI \cite{HEAI}&
  Europe & H2020 EU Project&
   To create the foundations for AI systems that empower people and society, with special focus on \textbf{Collaborative Humane Computer Interaction} based on a convergence of HCI with ML.\\ \hline  
AI Watch \cite{AIW}&
  European Commission & Public Institution&
   An initiative to monitor the development, uptake and impact of AI for Europe\\ \hline      
AI4People \cite{AI4People}&
  European Commission & Multi-stakeholder Forum&
  To \textbf{bring together all actors interested in shaping the social impact of new applications of AI}, including the European Parliament, civil society organisations, industry and the media. They published the AI4People's Ethical Framework \cite{FloridiAI4People} which inspired the TAIG \cite{HLEG2019}. \\ \hline
OECD.AI \cite{OECDAI}&
  Inter-governmental & International Organisation&
  The OECD AI Policy Observatory combines resources from across the OECD, its partners and all stakeholder groups to \textbf{facilitate dialogue between stakeholders} while providing \textbf{multidisciplinary, evidence-based policy analysis} in the areas where AI has the most impact. \\ \hline 
Knowledge Centre Data \& Society \cite{KCDS}&
  Belgium & Research Centre&
  Funded by the Flemish Department on Economy, Science and Innovation, it enables \textbf{socially responsible, ethical and legally appropriate implementations of AI in Flanders}.\\ \hline
DataEthics \cite{DataEth}&
  Denmark & ThinkDoTank&
  To ensure primacy of the human being in a world of data, based on a European legal and value-based framework. It has a core \textbf{focus on AI as the evolution of complex data processing} extended in human decision-making within politics, economics, identity and culture. \\ \hline
DATALAB - Center for Digital Social Research \cite{Datalab} &
  Denmark & Research Centre&
  Conducts research in many different aspects of behavioural data within several areas. A special focus is brought to the \textbf{social effects of automated data processing} as well as to the social adaptation of automated data systems.\\ \hline
ImpactAI \cite{ImpactAI}&
  France & Non-profit Association &
  Think\&Do Tank for Ethics and Responsible AI aiming to \textbf{promote the development of trusted AI}, support innovative projects and publish annual reports.\\ \hline
Algorithm Watch \cite{AlgWatch}&
  Germany & Non-profit Organisation &
   Based on research and advocacy to \textbf{evaluate algorithmic decision-making processes}, raise ethical conflicts and explain its features to general audience. \\ \hline
AI \& Society Lab \cite{AISL}&
  Germany & Research Laboratory&
  Interface and translator between academia on one side and industry and civil society on the other, it functions as \textbf{experimental space for new formats to advance knowledge generation and knowledge transfer to AI}.  \\ \hline
Institute for Ethics in AI \cite{IEAI} &
  Germany & Research Centre&
  To generate of global, egalitarian and interdisciplinary \textbf{guidelines} for the ethical development and implementation of AI and to integrate of \textbf{ethical and societal priorities into the development of fundamentally integrative AI} technologies.\\ \hline  
AI Sustainability Centre \cite{AISC}&
  Sweden & Consultancy&
  Creation of \textbf{AI Sustainability Framework} for identifying, measuring and governing the ethical implications of AI and assist organisations from a legal, technical and societal perspective. \\ \hline
AI Transparency Institute \cite{AITI}&
  Switzerland & Non-profit association & 
  Dedicated to \textbf{AI governance and human trust in AI}, they address key challenges in digital ethics, AI safety, transparency, fairness and privacy. \\ \hline
Digital Ethics Lab \cite{DEL}&
  UK & Research Centre &
  To tackle the \textbf{ethical challenges of digital innovation} from a multidisciplinary perspective, with the aim to identify benefits and positive opportunities while avoiding risks and shortcomings.\\ \hline
Institute for Ethical AI \& ML \cite{IEAIML}&
  UK & Research Centre &
  Highly-technical, practical and cross-functional research across \textbf{8 Machine Learning Principles} and \textbf{Explainable AI Framework}\\ \hline
Institute for Ethical AI in Education \cite{IEAIE}&
  UK & Research Centre&
  As a response to the TAIG, IEAIE works to \textbf{develop frameworks and mechanisms} to help ensure that the use of AI across education is designed and deployed ethically. \\ \hline
Leverhulme Centre for the Future of Intelligence \cite{LCIF} &
  UK & Research Centre&
  To build an interdisciplinary community of researchers with strong links to technologists and the policy world to study the \textbf{impact of AI in society with a focus on trust, fairness, accountability and democracy}.  \\ \hline
Centre for Data Ethics and Innovation \cite{CDEI} &
  UK & Public Institution&
  Part of Department for Digital, Culture, Media \& Sport, they connect policymakers, industry, civil society, and the public to \textbf{develop the right governance regime for data-driven technologies}. \\ \hline
\end{tabular}
\end{table*}

As we said, the creation of the Observatory takes place in a complex and dynamic context where an imprecise number of AI-related events populate the European calendar. Table \ref{tab:centres} includes a collection of European centres that are specifically dedicated to the research around AI and its impact on Society. These were selected form a larger set based on a search of simple keywords on Google engine (such as ``AI'', ``ethics'', and ``society''). 
The common aim among these institutions is to promote designs and developments of technologies that put upfront concepts such as social responsibility, trust or fairness. Some are dedicated to the creation of guides, others to define evaluation methods, but all have in common the will to create spaces for multidisciplinary dialogue. 



While the abundance of centres and projects dealing with AI and its social and ethical impact is a sign of cultural awareness and a source of knowledge, all these positive undertakings run the risk of isolation and self-referentiality. Therefore, OSAI should try to bridge this gap and promote cooperation and mutual knowledge. In addition, it will focus on areas that extend beyond the ethical and legal aspects, including also socio-economic and cultural elements (\eg how AI is perceived among European citizens, how the arts are presenting or using AI). 

The Observatory differs from these initiatives in several respects. In the first place, the OSAI focuses not only on articles and news, but also on people. Indeed, one of the motivating ideas behind the Observatory is the creation of a community of people who can contribute to the discussion of ELSEC-AI. Such a community can combine various types of subjects such as AI experts (\eg AI researchers and practitioners), specialists in any ELSEC-related field (ethicists, sociologists, lawyers, policy makers, artists, etc) and lay people. In the second place, the OSAI will approach ELSEC-AI in the context of Europe so as to foster the dialogue among European countries.

\section{OSAI Working Groups} \label{sec:WGs}
To revise how AI systems are evaluated and complement performance metrics with ELSEC-AI considerations, we need a change in the background of the process. A first step is to promote diversity in the teams that are both designing and assessing such systems. Note that when we talk about diversity, we do not only refer to gender, ethnicity or functional capabilities, but also to include professionals from multiple disciplines and domains of expertise. 


A second step is to foster a multidisciplinary dialogue among experts to promote reflection on ELSEC-AI and identify shared strategies to consider ELSEC issues within the AI life cycle.
OSAI is trying to fulfill these tasks (diversity and multidisciplinary discussion) by creating a set of working groups, \ie semi-organised groups of experts working on ELSEC-AI topics. 

The experience of the working groups is a laboratory to explore ways to address ELSEC-AI from a diversity of perspectives. They have an experimental character in that there is no systematic knowledge and expertise in dealing with ELSEC-AI in real-world with a multidisciplinary approach. 
At the beginning, working groups will emphasize two perspectives with a view to incorporating further aspects in future: 
\begin{itemize}
    \item Legal AI: to study existing laws and regulations, how applicable they are to AI systems and identify possible gaps.
    \item Ethical AI: to promote the design and development of AI systems that respect fundamental rights. 
\end{itemize}

The working groups will be formed by experts in different areas, from academic, business, media or other backgrounds: lawyers and data protection officers, philosophers, software engineers, journalists, sociologists, etc. With this variety of profiles we expect to generate a wide range of opinions and experiences around our topics of interest. Moreover, these working groups are supposed to grow with a bottom-up approach engaging participants from the very beginning. Participants will work in smaller groups and in a limited span time \footnote{Now, the schedule of activities will be constrained by the AI4EU lifetime, but this might be revised in future.}. This activity aims to encourage the participation of experts with an open and transparent methodology and engage them in the communication and sharing of knowledge in this new area of interest. The main objectives of this activity are the following:
\begin{itemize}
\item To fill the gap between the ethical debate and the engineering practice in European organizations.
\item To help researchers and practitioners navigate the ethical challenges that arise in different real-world AI applications.
\item To support interdisciplinary dialogue engaging people from different backgrounds.
\item To promote cross-fertilisation among different sectors (\eg academia, companies, public institutions).
\item To inspire future responsible practices in the field of AI.
\item To create ELSEC-AI literacy understandable for different types of audiences and domains of expertise.
\end{itemize}

To fulfill the last two points, we expect to generate a set of good AI practices with a focus on how to implement guidelines such as the HLEG-AI Trustworthy AI Guidelienes, especially for SMEs and start-ups. We will also adapt part of the content into educational material that will be shared through the Observatory and the AI4EU communication channels. 

\subsection{Background and Methodology}
In the second half of 2019, the authors participated in the piloting of the Trustworthy AI assessment List (TAIL) that was published along with the Trustworthy AI Guidelines by the HLEG-AI, were 50 companies from different sectors and European countries were interviewed. The result of that investigation suggested the need to promote operational resources (\eg tools, frameworks, procedures/methodologies, etc.) to support organisations to take the Trustworthy AI requirements into account. Other studies have stressed the value to share knowledge and experiences (\eg best practices, case studies) to help practitioners navigate complex ethical issues and interventions \cite{Holstein19}. 



To build upon the piloting of the TAIL, we propose to form small teams that can collaborate with companies (but also public organisations) that are implementing AI products with a view to experiment how to tackle Trustworthy AI requirements, also drawing on responsible practices (see section \ref{sec:RPs}), generate and share a list of good practices that can inspire other organisations. We hope in this way to help fill  the gap of knowledge and language between the different stakeholders, \ie from regulations, to recommendations to the final translation to software engineering methods and other sorts of processes (Human-Computer Interaction methods, ML approaches, management strategies, etc.). Also we will try to disseminate the result of working groups through the Observatory web-site so as to include the Society in this process of gaining trust in new technologies.


At present, the working groups are not intended to generate a comprehensive methodology or a new standard but identify good practices that help integrate ethical and legal requirements into the assessment of AI systems. In concrete, the teams of experts will interact with organisations to analyse specific case studies where they can apply and test one or more responsible AI practices. Interactions can take the form of interviews, design thinking sessions, algorithmic audits, among others, depending on the operational resources adopted by the team (see the responsible practices in Section \ref{sec:RPs}). This would help organisations to check whether their AI products meet Trustworthy AI requirements, and possibly re-frame their objectives and their Key Performance Indicator (KPIs) in the light of ethical and legal constraints. 

A series of training sessions with invited speakers (either external or internal to the working groups) will help the members of working groups to achieve a shared knowledge about responsible AI practices (questionnaires and checklists, frameworks, strategy guides and canvases, etc). Also, to test and refine our proposal, we plan to pilot working groups with an internal activity involving a few experts and partners (from research and industry) of the AI4EU consortium.


\subsection{Responsible Practices}\label{sec:RPs}
While many AI principles have been published, the translation of these into practices and processes is still at the very beginning. The main reason relies on their abstract nature, which makes them difficult for practitioners to operationalise. Even though awareness of the potential ethical issues is increasing at a fast rate, the AI community's
ability to take action to mitigate the associated risks is still at its infancy \cite{morley2019initial}. Operationalisation so far strongly depends on moral compass of individual since there are no legal binds to the existing ethical guidelines. Thus, a framework for ethical decision making and responsible practices is required.

We are classifying the responsible practices by their nature and intended use into 5 groups:
\begin{itemize}
    \item \textit{Assessments, questionnaires and checklists} raise questions, encourage reflection and inspire potential action, \eg the HLEG-AI TAIL with 131 questions to operationalise the seven key requirements declared in the AI guidelines \cite{HLEG2019}, or the Consequence Scanning, an agile practice to consider the potential consequences of a product or service on people, communities and the planet \cite{ConsScan}.
    \item \textit{End-to-end frameworks} address each stage of the entire process with appropriate activities and involve multiple audiences, \eg the End-to-End Framework for Internal Algorithmic Auditing to help companies and their engineering teams audit AI systems before deploying them \cite{accuntgap}, or the People + AI Guidebook to help user experience (UX) professionals and product managers follow a human-centered approach to AI \cite{PAIG}.
    \item \textit{Strategy guides and canvases} are thinking frameworks and diagnostic tools, that help break down and work through complex challenges, \eg the Data Ethics Canvas helps identify and manage data ethics considerations \cite{DEC}, or the Ethical Operating System to help inform the design and development process, provide strategies to mitigate risks and take action \cite{EthOS}.
    \item \textit{Design guides} are sets of recommendations towards responsible good practice in design, \eg the Guidelines for Human-AI Interaction recommend best practices for how AI systems should behave \cite{amershi2019guidelines}; AI Ethics Cards are a set of four design principles and ten activities that help guide an ethically responsible, culturally considerate, and humanistic approach to designing with data \cite{EthCom}.
    \item \textit{Software toolkits} provide metrics and algorithms to support the ethical development of AI-powered software, \eg the AI Fairness 360 Toolkit to help examine, report, and mitigate discrimination and bias in ML models throughout the AI application life cycle \cite{IBMbellamy}; Aequitas bias audit toolkit to audit machine learning models for discrimination and bias \cite{saleiro2018aequitas}.
\end{itemize}

\begin{figure}[h]
    \centering
    \includegraphics[width=\linewidth, height=5cm]{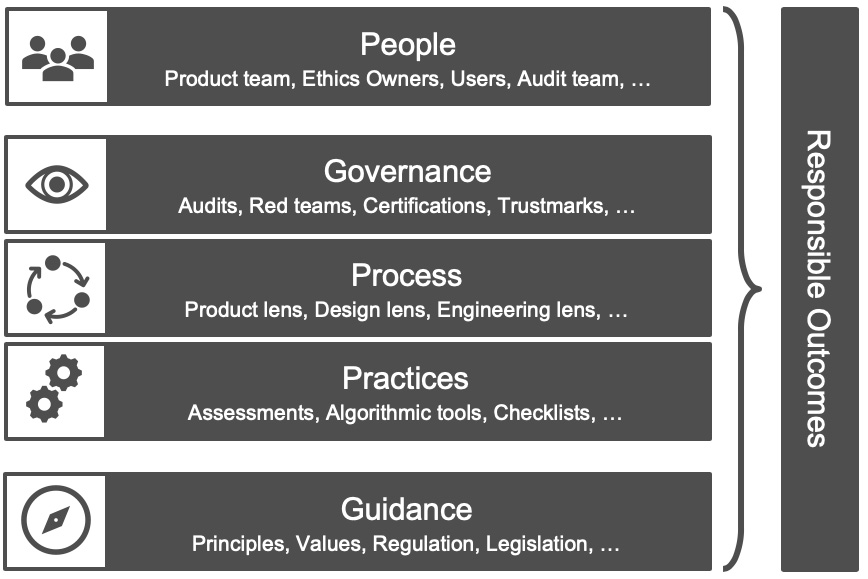}
    \caption{High-level overview of key pillars for progressing towards responsible outcomes}
    \label{fig:resp_prac}
\end{figure}

The landscape of responsible practices is wide, but insufficient, inefficient and scattered. Many of these practices do not adequately address the challenges of context-dependency, they lack ease of use \cite{AIEIG} and completeness, they either address only silo disciplines, or single process steps or particular problems. Furthermore, ticking boxes on fairness checklists, mitigating bias with algorithms and anticipating consequences with ethics cards is by far not enough. As depicted in Figure \ref{fig:resp_prac}, in order to progress towards responsible outcomes, it needs first and foremost close collaboration and a diverse range of perspective, guidance derived from values, principles and policies, a curated set of responsible practices, throughout the entire process, ensured by governance procedures, such as audit services, certifications and AI labels, or company-internal self-check tools, red teams, and ethics Objectives and Key Results (OKRs) and KPIs \cite{Metcalf2019OwningEC}. 

As companies start to envision procedures to operationalise AI principles, new professional roles, summarised as ``ethics owners'' by \cite{Metcalf2019OwningEC}, are being created in order to cover the lack of attention that ethics had in first place. They own responsibility for ethics practices across an organization, and engage to transform principles, values, ethical stances, and often legal and regulatory imperatives into concrete practices within their organization. To be most effective at achieving these goals, new responsible practices must be aligned with teams’ existing workflows and supported by organizational culture \cite{Madaio20}.

\section{Concluding remarks} \label{sec:conl}
This paper presents an overview on the existing initiatives that we can find today to enhance the progress of AI towards responsible design, development and deployment. This review puts special attention on the European centres and networks involved in this process since this is the main scope of the Observatory and the objectives within AI4EU. Although it is not exhaustive, it suggests that we are witnessing an \textit{ethical turn} promoting a wider discussion about the assessment of AI systems. In addition, the study on the different responsible practices shows an interest from a diversity of stakeholders to promote a different approach to the assessment of AI, although it is clear that there are still some gaps to cover in order to integrate the required ecosystem.


In his book Progress and its Problems, Larry Laudan recommends to cast the nets of appraisal sufficiently widely so as to include \emph{all} the cognitively relevant factors which are \emph{actually} present in the historical situation \cite{lau78}.
The OSAI aims to contribute to this change of paradigm by creating a space for multidisciplinary gathering, where different actors will be able to discuss and create literacy related to ELSEC-AI. We expect to study in depth the key pillars for progressing towards responsible outcomes of AI and their relation with the Trustworthy AI Guidelines and Assessment List in order to find existing relations, but also possible opportunities to complement them. Our final objective is to generate a set of good responsible practices that could help AI practitioners to implement the HLEG-AI documents and align with the European Commission's vision of Trustworthy AI. We strongly believe that this step is necessary to bring close all the stakeholders involved in the design, deployment and use of new technologies such as AI, which may have great benefits for the Society but are still in a process of being trusted.

\ack Teresa Scantamburlo and Atia Cortés are partially supported by the project A European AI On Demand Platform and Ecosystem (AI4EU) H2020-ICT-26 \#825619. The views expressed in this paper are not necessarily those of the consortium AI4EU.

\bibliography{main}
\end{document}